\documentclass[showkeys,showpacs]{revtex4}

\usepackage{eucal}
\usepackage{eufrak}
\usepackage{amsmath,amsfonts}
\usepackage{amssymb,latexsym}
\usepackage{color}
\usepackage{graphicx}

\newcommand{\ii}{\mathrm{i}}
\newcommand{\ud}{\mathrm{d}}
\newcommand{\uD}{\mathrm{D}}
\begin{document}

\title{Uncertainty relation for angle from a quantum-hydrodynamical perspective}
\author{J.-P.\ Gazeau}
\email{jeanpierregazeau@mac.com ,gazeau@apc.in2p3.fr }
\affiliation{Universit\'e de Paris, CNRS, Astroparticule et Cosmologie, F-75013 Paris, France}
\affiliation{Centro Brasileiro de Pesquisas F\'{i}sica, 22290-180, Rio de Janeiro, RJ, Brazil}
\author{T.\ Koide}
\email{tomoikoide@gmail.com,koide@if.ufrj.br}
\affiliation{Instituto de F\'{\i}sica, Universidade Federal do Rio de Janeiro, C.P. 68528,
21941-972, Rio de Janeiro, RJ, Brazil}

\begin{abstract}
We revisit the problem of the uncertainty relation for angle by using quantum hydrodynamics formulated in terms of
the stochastic variational method (SVM), where we  need not define the angle operator. 
We derive both  the Kennard and Robertson-Schr\"{o}dinger inequalities for canonical variables in polar coordinates.
The inequalities have state-dependent minimum values which can be smaller than $\hbar/2$ and then 
permit a finite uncertainty of angle for the eigenstate of the angular momentum. 
The present approach provides a useful methodology to study quantum behaviors in arbitrary coordinates.
\end{abstract}



\maketitle

\section{introduction}

The uncertainty relation is known as one of the important features of quantum physics and 
its comprehension requires unceasing improvement.
For example, its error-disturbance generalization proposed by Ozawa \cite{ozawa1} was experimentally tested \cite{ozawa3,ozawa2,ozawa4,ozawa5}. The error  quantifies the deviation between  intended and effective measurements, and the disturbance  concerns accuracy in the successive determinations of the value of two non-commuting observables. As a matter of fact, the concept of  uncertainty relation is crucial  for experimental requirements in the detection of the gravitational waves \cite{maddox88,cripe19}. 

In this work, we consider a long-standing problem, that is, the uncertainty relation for 
the angular coordinate viewed as the canonically conjugate variable to the angular momentum.
As is discussed in the standard textbooks of quantum mechanics \cite{book:schiff}, if one defines an angle operator 
$\theta$ 
such that  its commutation rule with the  angular momentum operator $L$ is canonical, i.e., $[\theta, L] = \ii\hbar$, then the ensuing Kennard inequality reads:  
\begin{eqnarray}
\sigma_\theta \sigma_L \ge \frac{\hbar}{2}\, ,
\end{eqnarray}
where $\sigma_A$ is the standard deviation of operator $A$.  
This is however not acceptable; the uncertainty of the angle $\sigma_\theta$ becomes infinite for the eigenstate of $L$, while 
the maximum value of $\sigma_\theta$ should be finite due to the bounded domain of the spectrum $0 \le \theta < 2\pi$.

This paradox is usually attributed to the definition of the angle operator: 
there is no self-adjoint multiplicative operator which is periodic and satisfies the canonical commutation rule.
Therefore the uncertainty relation for angle has been discussed exclusively by violating one of the two conditions. 
See Ref.\ \cite{review} and the recent Refs.\ \cite{kas06,jp1,diego18} for  survey and discussion. 
To avoid the problem of the periodicity, for example, the non-hermitian multiplicative operator with spectrum $e^{\ii\theta}$ is considered instead of the angular coordinate itself \cite{ohnuki,tanimura}. 
For the latter, the commutation rule is modified by 
\begin{eqnarray}
[\theta,  L] = \ii\hbar \left\{  1 - 2\pi \sum_{N=-\infty}^\infty \delta(\theta - 2N\pi) \right\}\, ,
\end{eqnarray}
where $N$ indicates an integer \cite{judge,judge2}. 

We investigate this problem from a different point of view which pertains to quantum hydrodynamics. 
Although such an approach gives an alternative point of view for quantum behaviors \cite{book:holland}, 
one could argue that quantum hydrodynamics does not offer an advantage over the  formalism of operators  acting in  Hilbert spaces of quantum states. 
However, with the help of the stochastic variational method (SVM) \cite{yasue,koide18}, 
quantum hydrodynamics provides a useful methodology to study quantum behaviors in arbitrary coordinates.
In fact, one of the present authors developed a systematic procedure to define the uncertainty relation 
within the framework of quantum hydrodynamics with Cartesian coordinates \cite{koide18}. 
Observables are described by using  probability distributions and  fluid velocity fields. Hence 
it is not necessary to work with position and momentum operators.
Moreover this framework was recently generalized to be applied to general coordinate systems in Ref.\ \cite{koide19}. 
Therefore the quantum-hydrodynamical approach is expected to cast the question of uncertain relation in a new light.
The purpose of this work is to examine the uncertainty relations for canonical variables in  polar coordinates by
using the quantum-hydrodynamical approach formulated in terms of the generalized SVM.

This paper is organized as follows. 
In Sec.\ \ref{sec:svm}, the formulation of SVM is discussed. Using this approach, we quantize a classical system in polar coordinates and the corresponding quantum hydrodynamics is obtained.
In Sec.\ \ref{sec:eigen},  eigenstates of angular momentum are discussed within the framework of SVM. The discussion in this section is not relevant to 
the derivation of the uncertainty relations.
The uncertainty relations in polar coordinates are derived in Sec.\ \ref{sec:uc}.
Section \ref{sec:conc} is devoted to concluding remarks.

\section{stochastic variation} \label{sec:svm}

The formulation developed here is the same as that in Refs.\ \cite{koide19,koide19-2}. 
We denote Cartesian coordinates by $z^{a} = (x,y)$ and polar coordinates by $q^i = (r,\theta)$.
Cartesian coordinates are  function of polar coordinates, $z^{a} = z^{a}(r,\theta) $.
In the following, $(\widehat{~~})$ represents a stochastic quantity and 
the Einstein notation of the summation is employed.
For the Euclidean SVM, see also review papers Refs.\  \cite{zam,koide-review}.

Before describing our formulation, let us give an outline of the SVM approach.
In classical mechanics, the Newton equation of motion is formulated through the variational principle applied to the Lagrangian. The latter  is 
defined in terms of kinetic part  $K$ and potential $V$ and is denoted $K-V$.
In this standard scheme of variation, the variations associated only with differentiable trajectories are considered.
In Ref.\ \cite{yasue}, by extending Nelson's stochastic mechanics \cite{nelson}, 
Yasue showed that the variation of the classical Lagrangian $K-V$ leads to the Schr\"{o}dinger equation in 
considering both  differential and non-differential trajectories in variation. 
This generalized theory of variation is called stochastic variational method (SVM) \cite{yasue,zam,koide-review,koide19,koide19-2,koide18,koide12}.
There are two important modifications in  applying the variational principle to the non-differential trajectory: 
one is the introduction of two Brownian motions  
in order to consider the non-differential trajectory in a time-reversal process (Eqs. (\ref{eqn:eq_r}) and (\ref{eqn:eq_r2})) 
and the other consists in generalizing  the time derivative in order  to apply it to
the non-differential (stochastic) trajectory (Eqs. (\ref{eqn:ftd}) and (\ref{eqn:btd})).

In SVM, we consider the optimization of the non-differentiable trajectory for a (virtual) particle. 
To describe such a trajectory as Brownian motion, we introduce stochastic differential equations (SDE's), which are obtained from Eqs.\ (2) and (7) in Ref.\ \cite{koide19} 
by substituting the metric (vielbein) of polar coordinates. In this approach, we introduce four unknown $u^i_\pm (r,\theta,t)$ $(i=r,\theta)$ to be determined through stochastic variational principle. These are requested to fulfill the periodicity conditions: 
\begin{eqnarray}
u^i_\pm (r,\theta,t) = u^i_\pm (r,\theta+2N\pi,t) \,, \ \mbox{for integer} \ N. \label{eqn:u-perio}
\end{eqnarray}
Then the evolutions forward in time are described by the forward SDE's ($dt >0$),
\begin{eqnarray}
\label{eqn:eq_r}
\begin{split}
\ud\widehat{r}_t 
&=  u^r_+ (\widehat{r}_t,\widehat{\theta}_t,t) \ud t +\frac{\nu}{\widehat{r}_t} \ud t  +  \sqrt{2\nu}  \ud\widehat{W}^r_t \, ,  \\
\ud\widehat{\theta}_t 
&=  u^\theta_+  (\widehat{r}_t,\widehat{\theta}_t,t)\ud t + \frac{\sqrt{2\nu}}{\widehat{r}_t} \ud\widehat{W}^\theta_t \, .
\end{split}
\end{eqnarray}
Here $\ud\widehat{A}_t = \widehat{A}_{t+\ud t} -\widehat{A}_t$ and $\nu$ is a parameter which characterizes the stochasticity. In the SVM application to quantum mechanics, $\nu$ is determined by the Planck constant $\hbar$, while it is described by the coefficient of viscosity in hydrodynamics \cite{koide12,koide18,koide19-2}.
The stochastic variables $\widehat{W}^r_t$ and $\widehat{W}^\theta_t$ 
are defined by 
\begin{eqnarray}
\label{wiener1}
\begin{split}
\ud\widehat{W}^r_t &\equiv \cos \widehat{\theta}_t  \, \ud\widehat{W}^x_t + \sin \widehat{\theta}_t  \, \ud\widehat{W}^y_t  \, , \\
\ud\widehat{W}^\theta_t &\equiv - \sin \widehat{\theta}_t  \, \ud\widehat{W}^x_t + \cos  \widehat{\theta}_t  \, \ud\widehat{W}^y_t\, ,
\end{split}
\end{eqnarray}
where, $\widehat{W}^{x}_t$ and $\widehat{W}^{y}_t$ describe the Wiener processes in Cartesian coordinates satisfying the standard correlation properties (a,b=x,y),
\begin{eqnarray}
E[\ud\widehat{W}^a_t] = 0\, , \quad  E[(\ud\widehat{W}^a_t)(\ud\widehat{W}^b_{t^\prime})] = \ud t\, \delta^{a b} \delta_{t,t^\prime} \,  . 
\end{eqnarray}
The stochastic ensemble average is denoted by $E[\quad ]$.
Note that the second term on the right-hand side of the radial component of Eq.\ (\ref{eqn:eq_r}) is induced by the correlations between $\widehat{W}^{x(y)}_t$ 
and the metric (vielbein) of polar coordinates, and 
prevents $\widehat{r}_t$ from being negative. 
See Refs.\ \cite{koide19,koide19-2,nagasawa}

In the variational method, we fix not only the initial but also the final distributions of particles. Then we need also to consider the 
evolutions backward in time, which are defined by the backward SDE's ($dt<0$),
\begin{eqnarray}
\label{eqn:eq_r2} 
\begin{split}
\ud\widehat{r}_t 
&=  u^r_- (\widehat{r}_t,\widehat{\theta}_t,t)\ud t - \frac{\nu}{\widehat{r}_t}  \ud t  +  \sqrt{2\nu}  \ud\underline{\widehat{W}}^r_{\;t}\, , \\
\ud\widehat{\theta}_t 
&=  u^\theta_-  (\widehat{r}_t,\widehat{\theta}_t,t)\ud t +  \frac{\sqrt{2\nu}}{\widehat{r}_t} \ud\underline{\widehat{W}}^\theta_{\;t} \, .
\end{split}
\end{eqnarray}
Here $ \ud\widehat{\underline{W}}^r_{\;t}$  and $\ud\underline{\widehat{W}}^\theta_{\;t}$ satisfy the same correlation properties as $\ud\widehat{W}^r_t$ 
and $\ud\widehat{W}^\theta_t$, respectively,  through replacing $\ud t$ with $|\ud t|$.

The particle probability distribution is defined by 
\begin{eqnarray}
\rho(r,\theta,t) &=& \frac{1}{r}  \int^\infty_0 r_i \ud r_i \int^{2\pi}_0   \ud\theta_i \, \rho_0 (r_i,\theta_i)
 \sum_{N=-\infty}^\infty E [\delta(r- \widehat{r}_t)\delta(\theta + 2N \pi - \widehat{\theta}_t )]\, , 
\end{eqnarray}
where the factor $1/r$ comes from the Jacobian and $\rho_0 (r_i,\theta_i)$ denotes the initial distribution of $\widehat{r}_t$ and $\widehat{\theta}_t$ at an initial time $t_i$.
It is easy to check the normalization $\int^\infty_0 r \ud r \int^{2\pi}_0 \ud\theta\, \rho(r,\theta,t) = 1$.

It is easily shown that the time evolution of $\rho$ satisfies two Fokker-Planck equations.
We do not make them explicit here (See Refs.\ \cite{koide19,koide19-2}), but just indicate that they derive from the two sets 
of SDE's \eqref{eqn:eq_r} and \eqref{eqn:eq_r2}, respectively. 
For these equations to describe the same phenomenon, the following consistency conditions should be satisfied, 
\begin{eqnarray}
& u^i_+ (r,\theta,t) - u^i_- (r,\theta,t) = 2\nu g^{ij} \partial_j \ln \rho (r,\theta,t)
\, , \label{eqn:cc}
\end{eqnarray}
where $i,j=r,\theta$ and $g^{ij}$ is the metric of polar coordinates.
This is an important property to derive the uncertainty relation. 
Then the two Fokker-Planck equations are reduced to the common equation of continuity,
\begin{eqnarray}
\partial_t \rho = - \nabla_i \left( \rho \frac{u^{i}_+  + u^{i}_- }{2} \right) \equiv  - \nabla_i (\rho v^{i})\, , \label{eqn:eoc}
\end{eqnarray}
where $\nabla_i$ is the covariant derivative in  polar coordinates.

The kinetic term of the classical Lagrangian has a time derivative term which should be replaced 
by a stochastic quantity.
This is however not trivial because the stochastic trajectory is non-differential and the standard definition of the time derivative is not applicable. 
Nelson introduced two time derivatives \cite{nelson}; 
the mean forward derivative,
\begin{equation}
\uD_+  f(\widehat{q}^{\,i}_t)  = \lim_{\ud t \rightarrow0+} E \left[  \frac{ f(\widehat{q}^{\, i}_{t + \ud t}) -
f(\widehat{q}^{\,i}_t)}{\ud t} \Big| \mathcal{P}_{t} \right]\,  , 
\label{eqn:ftd}
\end{equation}
and the mean backward derivative,
\begin{equation}
\uD_-  f(\widehat{q}^{\,i}_t)  = \lim_{\ud t \rightarrow0-} E \left[ \frac{ f(\widehat{q}^{\,i}_{t + \ud t}) -
f(\widehat{q}^{\,i}_t)}{\ud t} \Big| \mathcal{F}_{t} \right]\, .
\label{eqn:btd}
\end{equation}
These expectation values are conditional averages, where the condition $\mathcal{P}_{t}$ ($\mathcal{F}_{t}$) 
fixes the values of $\widehat{q}^{\, i}_{t^\prime}$ for
$t^{\prime}\le t~~(t^{\prime}\ge t)$. 
The trajectory becomes smooth and $\uD_+$ and $\uD_-$ coincide in the vanishing limit of $\nu$.

The action for the stochastic variation is defined by 
\begin{eqnarray}
I = \int^{t_f}_{t_i} \ud t E[ L ]\, ,
\end{eqnarray}
with the stochastic Lagrangian, 
\begin{eqnarray}
L
&=& 
\frac{m}{4}\sum_{s=\pm} \sum_{a=x,y}( \uD_s \widehat{z}_t^{\, a})\, ( \uD_s \widehat{z}_t^{\, a}) - V \, .
\end{eqnarray}
Here $m$ is the particle mass, $V$ is a potential energy and 
$\widehat{z}^{\, a}_t  = z^{a}(\widehat{r}_t, \widehat{\theta}_t)$.
We have replaced the classical kinetic term by the average of the contributions from $\uD_+$ and $\uD_-$. 
Note that the above stochastic Lagrangian reproduces the classical particle Lagrangian in  polar coordinates 
in the vanishing limit of the noise $\nu$.
See Refs. \cite{koide12,koide18,koide19-2} for other choices of the kinetic term in the stochastic Lagrangian.

We require that the above action be optimized, not only for any variation of the stochastic trajectory, but also for
any distribution of the stochastic polar coordinates. 
Then the stochastic variation of the action determines the equations for the velocity field $v^{i}(r,\theta,t)$ (or equivalently $u^{i}_\pm (r,\theta,t)$) 
and leads to the following quantum hydrodynamics, 
\begin{eqnarray}
\begin{split}
\left(\partial_t + v^r \partial_r + v^\theta \partial_\theta\right)v^r 
-r v^\theta v^\theta 
+  \frac{\partial_r V }{m} 
&= 
2\nu^2
\partial_r \frac{1}{\sqrt{\rho}} 
\left(
\frac{1}{r} \partial_r r \partial_r + \frac{1}{r} \partial^2_\theta
\right)
\sqrt{\rho}\,,\\
\left(\partial_t + v^r \partial_r + v^\theta \partial_\theta\right) v^\theta 
+ \frac{2}{r} v^r v^\theta 
+ \frac{\partial_\theta V }{mr^2} 
&=
\frac{2\nu^2} {r^2}\partial_\theta \frac{1}{\sqrt{\rho}} 
\left(
\frac{1}{r} \partial_r r \partial_r + \frac{1}{r} \partial^2_\theta
\right)
\sqrt{\rho} \, . 
\end{split}
\label{eqn:qhtheta}
\end{eqnarray}
All quantum effects are produced by the terms appearing in the respective right-hand sides of these two equations. 
They are given by the gradient of the so-called quantum potential \cite{book:holland}.

To see the relation to quantum mechanics, we introduce the velocity potential through
\begin{eqnarray}
v^{i} (r,\theta,t) = 2\nu g^{ij}\partial_j \Theta (r,\theta,t)\, , \label{eqn:phase}
\end{eqnarray}
and define the complex function $\Psi = \sqrt{\rho} e^{\ii \Theta}$.
By choosing $\nu = \hbar/(2m)$, one can easily show that $\Psi$ 
satisfies the Schr\"{o}dinger equation in  polar coordinates and thus $\Psi$ is viewed as a wave function. 
See also Eq.\ (48) in Ref. \cite{koide12}.
That is, this approach enables us to quantize directly classical systems represented in terms of polar coordinates.
This is a remarkable feature which does not held if one proceeds with the canonical quantization.

The quantum potential term become singular as $\sqrt{\rho}$ crosses zero, which corresponds to 
the node of the wave function.
Then we need an additional condition to connect the solutions around the singularity.
This is indeed associated with the criticism by Takabayasi and Wallstrom, as we will show below \cite{tw1,tw2} 
and described in the next section.

For the sake of simplicity  in the following discussions, we suppose that the potential is a function  of $r$, $V=V(r)$ only and put $\nu=\hbar/(2m)$. 
For the SVM formulation of hydrodynamics, see Refs.\ \cite{koide12,koide18,koide19-2} for details.

\section{Eigenstate of angular momentum} \label{sec:eigen}

Before deriving the uncertainty relation, we comment how the eigenstate of the angular momentum is described in this scheme.

The eigenstate is interpreted as the stationary solution of quantum hydrodynamics, 
\begin{eqnarray}
 v_{r}(r,\theta,t) = 0\, , \ \ \
 v_{\theta} (r,\theta,t) = \frac{\hbar \alpha}{m}\, , \label{eqn:sta-sol}
\end{eqnarray}
where $\alpha$ is still an arbitrary adimensional real constant. We consider the covariant component $v_i$, not the contravariant component $v^{i}$.
Substituting these into Eq.\ (\ref{eqn:eoc}) and into the second line of Eq.\ (\ref{eqn:qhtheta}), we find $\rho(r,\theta,t) = \rho(\theta)$.
Then the first line in Eq.\ (\ref{eqn:qhtheta}) becomes the time-independent Schr\"{o}dinger equation in  polar coordinates, 
\begin{eqnarray}
\left[- \frac{\hbar^2}{2m} 
\left(
\frac{1}{r} \partial_r r \partial_r 
-\frac{\alpha^2}{ r^2}
\right)
+
  V (r) \right]\sqrt{\rho}
= \varepsilon \sqrt{\rho}\, , \label{eqn:tindsch}
\end{eqnarray}
where $\varepsilon$ is a real constant, that is, the energy eigenvalue. 
The same equation can be derived from the Bernoulli theorem of quantum hydrodynamics.

Note however that $\alpha$ is not yet quantized.  In fact, the quantum potential (the right-hand side of Eqs. (\ref{eqn:qhtheta})) becomes singular at $\sqrt{\rho} =0$ 
which corresponds to the nodes of wave function 
and thus we need additional condition to connect the solutions for the left and right sides of the singularity. 
In quantum hydrodynamics, the standard procedure amounts to employ the Bohr-Sommerfeld type condition  \cite{tw1,tw2,book:holland},  
\begin{eqnarray}
\oint  v_\theta (r,\theta) \ud\theta = \frac{2\pi N \hbar}{m}\, ,
\end{eqnarray}
where the loop integral path is chosen to be around $r=0$. 
Then one can easily find the quantization of the angular momentum $\alpha=N$. 
See discussions in Ref.\ \cite{book:holland} for more details.

\section{Uncertainty relations} \label{sec:uc}

Following Ref.\ \cite{koide18}, we define the stochastic momenta by 
\begin{eqnarray}
\widehat{p}^{\,\pm}_{i \, t} &=& 2 \frac{\partial L}{\partial (\uD_\pm \widehat{q}^{\, i}_t)} =  m g_{ij} u^{j}_\pm (\widehat{r}_t,\widehat{\theta}_t,t)\, ,
\label{eqn:momentum2}
\end{eqnarray}
where, in the present case, $\widehat{q}^{\,r}_t = \widehat{r}_t$ and $\widehat{q}^{\,\theta}_t = \widehat{\theta}_t$.
The factor $2$ is just a convention to reproduce the classical result in the vanishing limit of $\hbar$ (or equivalently $\nu$). 
It is easy to show that the expectation value of the quantum-canonical momentum is given by the average of the two expectation values of the two momenta, 
\begin{eqnarray}
E[ \widehat{p}_{i\, t} ]
\equiv
E\left[\frac{\widehat{p}^{\,+}_{i \, t}  + \widehat{p}^{\,-}_{i \, t}}{2}\right] 
=
m \int^\infty_0 r \ud r \int^{2\pi}_0\, \ud\theta\, v_i (r,\theta,t) \rho(r,\theta,t)
\, . \label{eqn:1storder} 
\end{eqnarray}
On the other hand, 
the stochastic momenta $\widehat{p}^{\,+}_{i \, t}$ and $\widehat{p}^{\,-}_{i \, t}$ contribute 
to the stochastic Newton (or Newton-Nelson) equation on an equal footing as is shown in Ref.\ \cite{koide18} (see Eq.\ (23)) and thus 
it is natural to define the standard deviation of the quantum-mechanical momentum by the average of the two contributions,
\begin{eqnarray}
\sigma^2_{p_i} \equiv 
\frac{1}{2} \left( \Delta^2_{ p^{\,+}_i } + \Delta^2_{p^{\,-}_i} \right) \, , \label{eqn:sigma-p}
\end{eqnarray}
where $\Delta^2_{ A } =  E\left[ (\delta \widehat{A})^2 \right]$ with $\delta \widehat{A} = \widehat{A} - E[\widehat{A}]$.
Note that Eq.\ (\ref{eqn:sigma-p}) can be reexpressed as
\begin{eqnarray}
\sigma^2_{p_i} = \Delta^2_{(p^+_i - p^-_i)/2} + \Delta^2_{(p^+_i + p^-_i)/2} \, .
\end{eqnarray} 
Then, using the Cauchy–Schwarz inequality, it is easy to show for a spatially $D$-dimensional system that 
\begin{eqnarray}
\Delta^2_{q^{i}} \Delta^2_{(p^+_j + p^-_j)/2}
&\ge& 
\left|   E[ \delta \widehat{q}^{\, i}_t \, \delta \widehat{p}_{j\, t} ]  \right|^2 \, ,   \label{eqn:uncer-part1}\\
\Delta^2_{q^{i}} \Delta^2_{(p^+_j - p^-_j)/2}
&\ge& 
\frac{1}{4}\left|   E[ \widehat{q}^{\, i}_t (\widehat{p}^{\, +}_{j\, t} - \widehat{p}^{\, -}_{j\, t} )]  - E[ \widehat{q}^{\, i}_t ]  E[ \widehat{p}^{\, +}_{j\, t} - \widehat{p}^{\, -}_{j\, t}  ]\right|^2  
\nonumber \\
&=& 
\frac{\hbar^2}{4}\left|
 \delta^{i}_{j} 
- 
\int \ud^D q\, \partial_j \{ J \rho \left(q^{i} - E\left[\widehat{q}^{\,i}_t\right]\right)\}
+ \int J \ud^D q\, \rho\,  \left( q^{i} - E\left[\widehat{q}^{\,i}_t \right] \right) \Gamma^k_{jk} 
\right|^2  \, , \label{eqn:uncer-part2}
\end{eqnarray}
where $J$ is the Jacobian, $\Gamma^i_{jk}$ is the Christoffel symbol and $\widehat{p}_{j\, t}$ is defined by Eq.\ (\ref{eqn:1storder}).
In the derivation of Eq.\ (\ref{eqn:uncer-part2}), the consistency condition (\ref{eqn:cc}) is utilized.
For Cartesian coordinates, the right-hand side of Eq.\ (\ref{eqn:uncer-part2}) is simplified as $\frac{\hbar^2}{4} \delta^{i}_j$, which leads to 
the standard minimum value of the Cartesian uncertainty relation. 
The detailed derivations of the standard Kennard ans Robertson-Schr\"{o}dinger inequalities are given in Sec. 3.1 of Ref.\ \cite{koide18}.

Like for the uncertainty relation of the quantum-canonical variables, 
we consider the product of the standard deviation of the {\it contravariant} components of canonical coordinates and that of the {\it covariant} components of canonical momenta.
Let us first consider the radial component. 
Substituting $(q^{i}, p_j) = (r, p_r)$ and the metric $g_{ij} = {\rm diag} (1, r^2)$, the right-hand side of Eq.\ (\ref{eqn:uncer-part2}) is calculated. 
The inequality for the radial component is then obtained as
\begin{eqnarray}
\sigma^2_{r} \sigma^2_{p_{r}}
&=& \Delta^2_{r} \Delta^2_{(p^+_r - p^-_r)/2} + \Delta^2_{r} \Delta^2_{(p^+_r + p^-_r)/2} \nonumber \\
&\ge& \frac{\hbar^2}{4} 
\left|
2-
E\left[ \widehat{r}_t \right]E\left[\widehat{r}^{\, -1}_t \right]
 \right|^2 
 + 
\left| E\left[ \delta \widehat{r}_t  \,
\delta \widehat{p}_{r\, t}
\right] \right|^2 \, ,\label{eqn:kennerd-r1}
\end{eqnarray}
where $\sigma^2_{r}  \equiv \Delta^2_{r} $.
This represents the Robertson-Schr\"{o}dinger inequality for $r$. 
When the second term of the second line is dropped, 
it becomes the corresponding Kennard inequality. 
Because 
$E[\widehat{r}_t]E[\widehat{r}^{\, -1}_t] \ge 1$, 
the right-hand side can be smaller than the standard minimum value of the uncertainty.

Similarly the inequality for the angular component is 
\begin{eqnarray}
\sigma^2_{\theta} \sigma^2_{p_{\theta}}
&=& \Delta^2_{\theta} \Delta^2_{(p^+_\theta - p^-_\theta)/2} + \Delta^2_{\theta} \Delta^2_{(p^+_\theta + p^-_\theta)/2} 
\nonumber \\
&\ge&
\frac{\hbar^2}{4} 
\left|
1
- 2\pi \int ^\infty_0 r \ud r  \rho(r,2\pi,t) 
 \right|^2 
 + 
\left| E\left[ \delta \widehat{\theta}_t  \, 
\delta \widehat{p}_{\theta\, t} 
\right] \right|^2 \, ,  
\label{eqn:kennerd-theta1}
\end{eqnarray}
where $\sigma^2_{\theta}  \equiv \Delta^2_{\theta} $.
This is the Robertson-Schr\"{o}dinger inequality for $\theta$ which reduces to the corresponding Kennard inequality by ignoring the second term of the second line.
For the eigenstate of the angular momentum, the distribution is homogeneous in the angle variable, $\int ^\infty_0 r \ud r \rho(r,2\pi,t) = 1/(2\pi)$, 
and then the right-hand side vanishes,
$\sigma^2_{\theta} \sigma^2_{p_{\theta}} \ge 0$.
That is, the standard deviation of the angle can be finite even for the eigenstate of the angular momentum, as is expected.

\section{Concluding remarks} \label{sec:conc}

Interestingly,  the well-known minimum of the uncertainty $\hbar/2$ is not 
a universal value for arbitrary canonical variables.
This fact may be useful to improve the resolutions of experiments on the microscopic level.

Judge considered a wave function on a circle and derived an uncertainty relation for angle which 
is the same as our Kennard inequality by ignoring the radial distribution and redefining $\sigma_{\theta}$ appropriately \cite{judge,judge2,review}.
In this derivation, the commutation rule is modified as was mentioned in the introduction.
Unfortunately the systematic procedure of such a modification is not known for arbitrary canonical variables.

On the other hand, the present derivation of the uncertainty relation is easily generalized to arbitrary coordinates in a non-relativistic systems. 
In fact, the most general representation of the Robertson-Schr\"{o}dinger inequality in the spatial $D$-dimensional system is 
easily obtained using Eqs.\ (\ref{eqn:uncer-part1}) and (\ref{eqn:uncer-part2}) as 
\begin{eqnarray}
\sigma^2_{q^{i}} \sigma^2_{p_{j}}
\ge
\frac{\hbar^2}{4}
\left| 
\delta^{i}_j
- \int \ud^D q\, \partial_j \{ J \rho \left(q^{i} - E\left[\widehat{q}^{\,i}_t\right]\right)\}
+  \int J \ud^D q\, \rho\,  \left( q^{i} - E\left[\widehat{q}^{\,i}_t \right] \right) \Gamma^k_{jk} 
 \right|^2 
+ 
\left| E\left[ \left(\delta \widehat{q}^{\,i}_t \right) \left(\delta \widehat{p}_{j\, t} \right) \right] \right|^2
  \, . \label{eqn:ucr-gene}
\end{eqnarray}
The Kennard inequality is obtained by dropping the second term on the right-hand side.
The term next to $\delta^{i}_j$ on the right-hand side gives a finite contribution for periodic variables as is the case of Eq.\ (\ref{eqn:kennerd-theta1}), 
and the term including the Christoffel symbol can reduce the standard minimum value of uncertainty as is seen in Eq.\ (\ref{eqn:kennerd-r1}).
One can easily confirm that the above inequality reproduces the uncertainty relations in Cartesian coordinates.

One can see that the minimum uncertainty for $\theta$ is attained by the eigenstate of the angular momentum because 
$g_{\theta\theta}u_{+}^\theta = g_{\theta\theta} u_{-}^\theta = v_\theta = \hbar N/m$.
On the other hand, even if $v_r$ is given by a constant, 
$u_{+}^r$ does not coincide with $u_-^r$ in general and hence 
the discussion of the corresponding state for $r$ is more difficult. 
The minimum uncertainty state in Eq.\ (\ref{eqn:ucr-gene}) is an open question.
This question will be related to study the coherent state 
and the emergence of classical behaviors from the quantum-hydrodynamical perspective \cite{dey, mathew, gazeau-book}.

The present quantum-hydrodynamical approach and the standard operator formulation are complementary.
Our approach provides a systematic procedure to find quantum behaviors in general coordinates 
while the operator formulation is based on a more tractable equation, the Schr\"{o}dinger equation, 
within the framework of Hilbertian functional analysis. 
Therefore, the studies on quantum hydrodynamics will contribute also to improve our understanding for the operator formulation of quantum mechanics.

Strictly speaking, the definitions of the canonical momenta in Eq.\  (\ref{eqn:momentum2}) 
are fully justified when the Hamiltonian formalism is constructed in stochastic analytical mechanics.
Moreover, the uncertainty relation can be introduced to viscous fluids in curved geometries, as was done for Euclidean hydrodynamics \cite{koide18}.
These generalizations are possible and will be reported in a forthcoming paper \cite{koide19-2,koide-forth}.

\vspace{0.5cm}

The authors thank T.\ Kodama and D.\ Noguera for useful discussions and comments.
T. K. acknowledges the financial support by CNPq (303468/2018-1).
A part of the work was developed under the project INCT-FNA Proc.\ No.\ 464898/2014-5. J.-P. G. thanks the Centro Brasileiro de Pesquisas F\'{i}sica for hospitality 
and the Programa de Capacita\c{c}\~ao Institucional - PCI for financial support.

\end{document}